\documentclass[conference]{IEEEtran}
\IEEEoverridecommandlockouts
\usepackage{cite}
\usepackage{amsmath,amssymb,amsfonts}
\usepackage{algorithmic}
\usepackage{graphicx}
\usepackage{caption, subcaption}
\usepackage{textcomp}
\usepackage{xcolor}

\def\BibTeX{{\rm B\kern-.05em{\sc i\kern-.025em b}\kern-.08em
    T\kern-.1667em\lower.7ex\hbox{E}\kern-.125emX}}
\begin{document}

\title{GANtron: Emotional Speech Synthesis with Generative Adversarial Networks}


 \author{\IEEEauthorblockN{Enrique Hortal}
 \IEEEauthorblockA{\textit{Department of Data Science and Knowledge Engineering} \\
 \textit{Maastricht University}\\
 Maastricht, The Netherlands. \\
 email: enrique.hortal@maastrichtuniversity.nl.}
  \and
 \IEEEauthorblockN{Rodrigo Brechard Alarcia}
 \IEEEauthorblockA{\textit{Department of Data Science and Knowledge Engineering} \\
 \textit{Maastricht University}\\
 Maastricht, The Netherlands.}
 }

\maketitle

\begin{abstract}
 Speech synthesis is used in a wide variety of industries. Nonetheless, it always sounds flat or robotic. The state of the art methods that allow for prosody control are very cumbersome to use and do not allow easy tuning. To tackle some of these drawbacks, in this work we target the implementation of a text-to-speech model where the inferred speech can be tuned with the desired emotions. To do so, we use Generative Adversarial Networks (GANs) together with a sequence-to-sequence model using an attention mechanism. We evaluate four different configurations considering different inputs and training strategies, study them and prove how our best model can generate speech files that lie in the same distribution as the initial training dataset. Additionally, a new strategy to boost the training convergence by applying a guided attention loss is proposed. 
\end{abstract}


\section{Introduction}
Speech synthesis consists of creating an artificial voice through algorithms. This technology can be used in a wide range of cases like voice assistants, audio-books, tutorials, translation, making disabled people's lives easier, etc. At the beginning, speech synthesis consisted of simply recording someone's voice, speaking thousands of words and stitching them together on-demand later on. Although this method behaves and it has been extensively used, the results are not optimal.

One of the problems that arises when dealing with speech synthesis is that the results tend to sound ``robotic''. The generated voices are flat and do not contain emotional content. Fortunately, speech synthesis is a prominent area of research and the rise of deep learning has made possible huge progress. However, the current state of the art methods are not faultless either. We can differentiate two main drawbacks, the computational needs \cite{tacotron2} and the control over the generated voice. 

Current state of the art end-to-end methods for speech synthesis use a two-stage approach where the first model takes raw text as input and generates a spectrogram and the second one (so-called vocoder) transforms the spectrogram into an audio waveform. In this paper we focus only on the text-to-spectrogram area of speech synthesis. More specifically we will create a text-to-Mel spectrogram model with an easy prosody control system using deep learning. To do so, we study several available datasets and we investigate how the use of labelled data can improve the results. We modify a state of the art model to accept labels (emotions) and it is also combined with adversarial training to improve the quality of the model. To that end, in this paper we explore the use of Generative Adversarial Networks (GANs) \cite{gan} to add emotions as input of text-to-speech (TTS) models. 

However, defining what emotions are is a difficult task since it is fluid in meaning \cite{describing-emotional-states}. The Oxford dictionary defines it as ``a strong feeling such as love or anger, or strong feelings in general'', a very general description. This also excludes key areas that are important when emotions are related to speech like arousal or attitude. Theoretically, it makes sense to say that full-blown emotions cannot be found on speech since pure emotions make people speechless or incoherent \cite{describing-emotional-states} \cite{science-of-emotion}. There is much controversy when trying to define what emotions actually are \cite{emotion-regulation}. Furthermore, applying the same stimulus to different individuals is not guaranteed to evoke the same response \cite{emotion-elicitation}. As a compromised, in this work we implemented a model which considers five different emotions (namely, anger, disgust, fear, happiness, sadness and neutral) with different intensity levels.  

In essence, the work presented in this paper includes three core novelties. First of all, as above-mentioned, a new model combining a text-to-Mel spectrogram model with an adversarial training approach is proposed. Secondly, this model is implemented taking the emotional content into consideration. To that end, the architecture incorporates emotion-related cues as input. Finally, with the aim of enhancing the training convergence of our architecture, a novel training strategy, based on guided attention loss, is introduced.

The remainder of this paper is structured as follows. Firstly, we explore the related work in Section \ref{sec:related-work}. Section \ref{sec:methodology} explains the methodology and datasets used in this work. Then, in Section \ref{sec:experiments} the experiment conducted are described and their results are summarized in Section \ref{sec:results}. Finally, we include the conclusions in Section \ref{sec:conclusions}.

\section{Related work}
\label{sec:related-work}
Before the deep learning revolution we are living right now, TTS systems consisted of very complicated language models that relied on huge datasets and human knowledge of the language. These models could be divided into two sections, a linguistic processor and a waveform generator. The linguistic processor takes raw text as input and generates features like phonetic strings, syllabification, and prosody. This part is a mix of human-created resources (text normalization, pronunciation dictionaries) and computer-based ones (to deal with the pronunciation of words not found in the dictionaries). The waveform generator receives the information extracted by the linguistic processor and generates the end raw audio waveform \cite{tts-decade-progress}. 

In that regard, WaveNet \cite{wavenet} was a breakthrough on audio synthesis thanks to its deep learning approach. It is able to generate speech from linguistic features and even model music. The main drawback of this model is that it is very slow. Nonetheless, it sparked the deep learning revolution in the vocoder's field. Parallel WaveNet \cite{oord2017parallelwavenet} improved WaveNet by adding parallel feed-forward processing and using probability density distillation. The idea of probability density distillation is similar to GANs. In this case, instead of a discriminator, there is a pre-trained WaveNet model acting as a teacher and a Parallel WaveNet acting as the student. The latter is trained to replicate the output of the teacher. This translates into a faster model without losing significant inference quality. In a similar vein, Clarinet \cite{ping2018clarinet}, a fully convolutional network using deep voice 3 as the encoder-decoder structure, becomes the first text-to-wave model. WaveGLOW \cite{prenger2018waveglow} is a flow-based model that combines the techniques from GLOW \cite{kingma2018glow} and WaveNet. It is as fast as Clarinet but with a simpler architecture, that makes it easier to train and to replicate the results. Parallel WaveGAN \cite{yamamoto2019parallelwavegan} uses Generative Adversarial Networks (GANs) to train the model. Compared to Clarinet, Parallel WaveGAN training and inference are 4.82 and 1.96 times faster respectively. Lastly, MelGAN \cite{kumar2019melgan}, a fully convolutional model that generalizes to unseen speakers, runs more than twice faster than real-time on CPU, and is 10 times faster than WaveGLOW. It does not use any noise as input to the generator and uses weight normalization. It also includes three different discriminators at different scales.

Deep learning also arrived to the text-to-spectrogram domain to improve quality and training speed. There are two main models families, based on Deep Voice \cite{deepvoice} and based on Tacotron \cite{tacotron}. The first version of Deep Voice follows similar architecture to traditional models but replaces every part with neural networks. It consists of five blocks: grapheme-to-phoneme model, segmentation model, phoneme duration model, fundamental frequency model, and audio synthesis model. Deep Voice 2 \cite{deepvoice2} introduced trainable speaker embeddings to the model and proves that a single neural TTS system can learn hundreds of unique voices from less than half an hour of data per speaker. Even if Deep Voice 3 \cite{deepvoice3} keeps the name, it is a completely new architecture, a sequence-to-sequence (seq2seq) fully convolutional model with an attention mechanism. 

For its part, Tacotron is a seq2seq model with an attention mechanism since its first version. It transforms raw text into a linear-scale spectrogram and applies the Griffin-Lim reconstruction to output audio waveforms. Tacotron 2 \cite{tacotron2} improves the previous version using WaveNet as vocoder instead of Griffin-Lim, Mean Square Error (MSE) for output loss instead of log-likelihood and adding a ``stop token'' layer to predict when the output sequence is completed. It also predicts Mel spectrograms (explained in Section \ref{sec:methodology}) instead of linear-scale spectrograms. Having both systems separated (seq2seq + WaveNet) allows for independent training, as long as both are trained in Mel-spectrograms sharing the same characteristics.

Very soon, researchers wanted to have more control over the synthesized voice (prosody control \cite{prosody-effect}). This attempt started by adding control to select different speaker's voices, age, and gender \cite{huang2021}. To that end, ``style tokens'' were introduced in Tacotron. They are latent variables that capture prosodic information not present in the text, while no data annotation was required, neither global nor local. This style tokens can be obtained in different manners: with context-based RNN attention \cite{latent-style-factors}, with a variational autoencoder \cite{vae} \cite{leglaive2020}, with the attention mechanism \cite{attention-all-you-need} (either self-attention \cite{yang2020} or multi-head attention \cite{style-tokens}) or with embedding like the ``Capacitron'' \cite{embed-cap-exp}. Recently, very impressive results have been shown in the use of unlabelled data, such as making a model sing when it has never seen anything similar, like Mellotron \cite{mellotron} or a prosody control of very high quality like Flowtron \cite{flowtron}.

\section{Methodology}
\label{sec:methodology}
Generative Adversarial Networks (GANs) have shown impressive results in the domain of content generation, while Tacotron 2 has proven to be very effective in creating human-like speech synthesis. Therefore, in this work, we combine Tacotron 2 with GANs to create a new model, GANtron.
GANtron is composed of a generator, a sequence-to-sequence model with an encoder-decoder structure, and a discriminator. The encoder receives the input text and transforms it into a latent space. The decoder translates the output of the encoder into a mel-spectrogram using the attention mechanism. It also accepts labels as input either in the encoder or in the decoder. The discriminator is trained to differentiate between real spectrograms and those created by the generator.

\subsection{Mel spectrogram}
A spectrogram is a representation of the frequencies that are present in a signal and how it varies over time. The problem with spectrograms is that they do not take into consideration human hearing capabilities. On the Hertz scale, two sounds that have the same distance, for example sounds with frequencies 100Hz and 600Hz or 10,000Hz and 10,500Hz, do not sound equally different to humans. This phenomenon is corrected with the Mel spectrogram. To that end, the Mel scale partitions the Hertz scale into bins 
and transforms each bin into a corresponding bin in the Mel scale. Hence, a Mel spectrogram is simply a spectrogram where the Y-axis is the Mel scale instead of the Hertz scale.


\begin{figure}
    \centering
    \includegraphics[width=\linewidth]{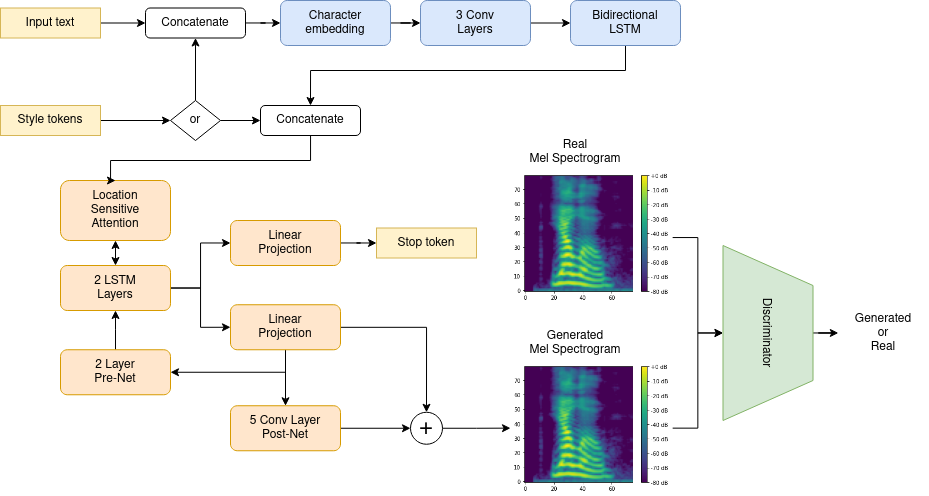}
    \caption{GANtron architecture.}
    \label{fig:GANtron}
\end{figure}

\subsection{Tacotron 2}
The paper describing Tacotron 2 divides it into two models, a modification of Tacotron and a modification of WaveNet \cite{tacotron2}. Tacotron is a seq2seq feature prediction network where the input is text and the prediction is a Mel-spectrogram. Wavenet translates the spectogram to waveforms. WaveNet is out of the scope of this work, and therefore, when we mention Tacotron 2 further in this paper, we only refer to the modification of Tacotron (the seq2seq model).

The encoder takes the input text as a list of characters and learns a character embedding. The embedding is sent to a stack of convolutional layers that model short-term context. Finally, it uses a bidirectional Long-Short Term Memory (LSTM) to generate the hidden encoded feature representations. Then, the decoder predicts the spectrogram one frame at a time, using the previous time step that is first processed by a Pre-Net. The outcome of the LSTM layers is sent to two different linear projection layers. The first is charged with predicting a ``stop token'' that indicates the end of the generation and the second predicts the Mel-spectrogram. To improve the result, the predicted Mel-spectrogram is processed by five convolutional layers as a residual Post-Net. 


\subsection{GANtron} 
In this work, we propose a novel approach, trying to combine the strenght of two state of the art approaches. In the introduced model, we expand Tacotron 2 to take advantage of its power to extract meaningful features of seq2seq nature. With that in mind, we apply generative adversarial networks, well-known for its ability to generate realistic samples in an unsupervised fashion. Therefore, the initial seq2seq architecture from Tacotron 2 is modified to include a Generator and a Discriminator that compete with each other during the training process (see Figure \ref{fig:GANtron}). 

It is worth noting that Tacotron 2 was trained for 500 thousand steps and, in our experiments with a Tesla P100-16GB, training for two days achieved around 65 thousand steps. This constraint creates a big computational challenge and the results achieved will probably have a reduced quality when compared to the original Tacotron 2. To mitigate this issue by increasing the speed at which the models' attention mechanism would converge, we propose the integration of a guided attention loss \cite{guidedattn}. 

Finally, the last novelty of the proposed approach is the introduction of emotional content in the text-to-speech architecture. For this purpose, we also investigate the use of datasets containing emotional content. Finally, this information is considered at different levels with the aim of exploring the effect of this emotional content in this framework, either using labels representing the emotions or not (refer to Section \ref{sec:trained_models} for further information).

\subsubsection{Generator}
In the case of text-to-speech, it is not possible to have a generator that only uses noise as input. In our architecture, the generator is a modification of the seq2seq model of Tacotron 2, adapted to have extra input conditions, called ``style tokens''. They are n-dimensional vectors used in concatenation with the input. When multiple speakers are used, then the speaker ID is the first value of the style token. The other values, depending on the experiment, are either random noise or labels representing emotions, in both cases values between 0 and 1 (see Section \ref{sec:trained_models}). Depending on which version, the number of parameters differs, ranging \~28M (only labels) to \~32M (labels and noise of size 512). GANtron has been designed so that the style tokens can be used either in the encoder or in the decoder. In the former, they are concatenated to the output of the character embedding layer. In the latter, they are concatenated to the output of the encoder.

\subsubsection{Discriminator}
\label{sec:discriminator}
The discriminator is a key part of the GAN architecture, it competes with the Generator, trying to differentiate between real and fake Mel-spectrograms. To improve the learning stability and to avoid mode collapse, WGAN \cite{wasserstein} has been implemented for the loss calculation. Furthermore, we have developed two different discriminators to investigate which one gives better results.

The first discriminator is composed of 1D convolutional layers. This allows processing the whole spectrogram in one go, no matter its size. To decide if a spectrogram is real or not, we take the average of the output for each frame. It has been designed in such a way that multiple frames can be packed together (windows). This discriminator will be referred to as ``convolutional discriminator''. This model consists of 4 $Conv1d+Dropout(0.5)+Tanh$ blocks, and a fifth one with only a $Conv1d$. The dimension of the first layer is calculated as: $window\_size*n\_mels$. In our case it would be $20*80=1600$, then reduced to 1024. The other blocks use 512 nodes and the last one is reduced to 80. The total number of parameters for this architecture is \~12M.

The second discriminator is composed of linear dense layers. For this reason, the input must be of fixed size. We process the whole spectrogram with a moving window, allowing overlapping between windows in a random fashion, with a maximum value allowed (notice that this randomness is not present in the first approach). To decide whether a spectrogram is real or not, the same procedure applies, taking the average of every processed window. We refer to this discriminator as ``linear discriminator''. In this case, we have 3 blocks, going the fourth layer from 512 (instead of 80) to 1. The total number of parameters for this architecture is \~1M.

\subsubsection{Loss function}
The loss used in this work is a combination of several loss functions performed as follows:
\begin{equation}
G_{loss} = Mel_{loss}  + Gate_{loss} + Wasserstein_{loss}  + Attn_{loss}
\end{equation}
where:
\begin{equation}
\begin{split}
Mel_{loss}  = MSE(mel_{(postnet-output)}, mel_{target}) \\ + MSE(mel_{output}, mel_{target})
\end{split}
\end{equation}

\begin{equation}
Gate_{loss} = BCE(gate_{out}, gate_{target})   
\end{equation}

Regarding the guided attention loss ($Attn_{loss}$), we use the one proposed in \cite{guidedattn}.

\subsubsection{Generator-discriminator training}
In connection with the iterative training of the generator and discriminator, it is noteworthy that finding a point of equilibrium between them is one of the challenges of the GANs training process. One of the critical points is to select a proper training ratio between the two networks. To choose the best discriminator-generator update ratio, a hyper-parameter search was run. The results of our experiments showed that training the generator twice after every discriminator update led the best results in our approach.


\subsection{Datasets}
\label{sec:datasets}
In this work, for training purposes, two datasets are used, each one with a different purpose. Firstly, a dataset intended to train the speech generation compose of several recorded hours is considered. This dataset, the LJ Speech \cite{ljspeech17} was also used by Tacotron 2 as it only needs to have audio files and their corresponding text. Secondly, a datasets labelled with emotions apart from the related text is needed. To that end, we considered VESUS \cite{VESUS}. This dataset has more audio files and more speakers and different utterances than similar options such as CREMA-D \cite{CREMA-D}, IEMOCAP \cite{IEMOCAP}, RAVDESS \cite{RAVDESS} or SAVEE \cite{SAVEE}. Table \ref{tab:datasets} includes a summary of the main characteristics of these datasets. Moreover, VESUS' authors did a phonetic comparison of the script and showed that their dataset is well-balanced. It has labels for each file given by annotators. After the normalization we concluded that this is the dataset with more emotional charge and the most balanced and therefore, it was chosen for the emotional factor of our model. 

Additionally, two extra datasets, namely CREMA-D and RAVDESS, are used for evaluation purposes. These dataset are concatenated with VESUS samples in order to investigate whether the generated samples contain emotional content (see Section \ref{sec:comparison} for further details).

\begin{table}[]
\centering
\caption{Publicly available datasets}
\label{tab:datasets}
\begin{tabular}{l|c|c|c|c|c}
\multicolumn{1}{c|}{Dataset} & \#speakers & \#unique  & \#files & \#emotions & Labels? \\ 
\multicolumn{1}{c|}{} &  & utterances &  &  &  \\
\hline
CREMA-D                      & 91         & 12                  & 7,442   & 6          & Y       \\
IEMOCAP                      & 10         & 8,068               & 10,039  & 9          & Y       \\
RAVDESS                      & 24         & 2                   & 1,044   & 8          & N       \\
SAVEE                        & 4          & 87                  & 480     & 7          & N       \\
VESUS                        & 10         & 253                 & 12,593  & 5 (+1)     & Y      
\end{tabular}
\end{table}

\begin{figure}[h]
    \begin{subfigure}[b]{0.48\textwidth}
      \includegraphics[width=\textwidth]{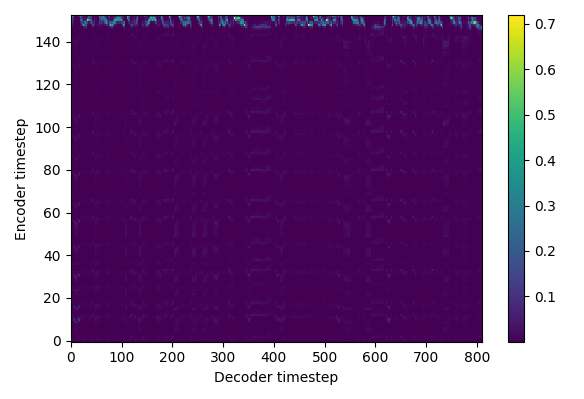}
      \caption{
      }
      \label{fig:no-warm-up}
    \end{subfigure}
    \begin{subfigure}[b]{0.48\textwidth}
      \centering
      \includegraphics[width=\linewidth ]{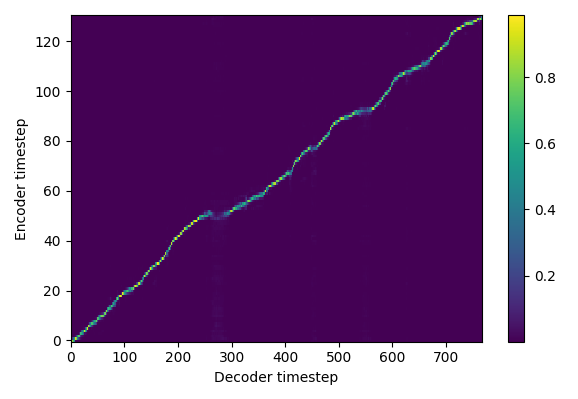}
      \caption{
      }
      \label{fig:warm-up-10k}
    \end{subfigure}
    \caption{Attention alignment at 30k steps, without guided attention warm-up (a) and at 10k steps, with guided attention warm-up stopped at 5k steps (b).}
    \label{fig:alignment}
\end{figure}

\subsection{Trained models}
\label{sec:trained_models}
 

Before the explanation of the models, some information regarding the style tokens is needed. To evaluate the importance of the input parameters and the training procedure, in this work we defined three alternative style tokens. Firstly, only noise is established as the most basic input to be used. It does not consider any emotional content. Secondly, a contextual style tokens is defined. In this case, as we are using, as labelled data, the VESUS dataset (see Section \ref{sec:datasets}), the style token is defined as a vector containing six values. Five of these values are the possible emotions considered in the dataset (namely, anger, fear, happiness, sadness and neutral). Additionally, whenever a model uses a dataset (or combination of datasets) that have more than one speaker, we need to include the speaker ID. There are two reasons for the inclusion of this information. Firstly to avoid introducing noise. If the model does not have the information of which speaker the spectrograms corresponds to, it will try to learn a voice that averages all of them. Secondly, to add more prosody control, being able to generate different voices with the same model. This speaker ID then becomes an extra value of the style token, increasing the initial number of tokens (emotions) by one. Finally, a combination of both, noise and contextual labels is considered as style token.


Based on the style tokens previously described and the applied training, four configurations were evaluated:
\begin{itemize}
    \item \textbf{Baseline model}: following the classical GAN architecture, we use only noise as style token. It is trained using only the LJ Speech dataset (refer to Section \ref{sec:datasets} for further details) with two variants, the style token being either as input to the encoder or to the decoder.
    \item \textbf{Expressive baseline}: this model is similar to the previous one but in this case, the VESUS dataset is included as part of the training set. However, in this model, the labels are not used but, since VESUS has more emotional content, we hypothesize that it can help the model to better grasp emotions.
    \item \textbf{Labelled model}: as in MelGAN, we created a GAN architecture where only the condition, without noise, is utilized as input. In this case, we use LJ Speech along with a labelled dataset (again, VESUS). Since LJ Speech has no labels, it was decided to assign a vector of zeros as contextual information.
    \item \textbf{Complete model}: the last variation combines the previous two models, using LJ Speech and VESUS as datasets as well but in this case, considering labels and noise as style token.
\end{itemize}


\section{Experiments}
\label{sec:experiments}
In this section, the experiments conducted to evaluate the four models presented in Section \ref{sec:trained_models} are described.

\subsection{Data preparation and visualization information}
To train and evaluate the models, we split each dataset randomly into train, evaluation, and test sets, with proportions 85\%, 5\% and 10\% respectively. Additionally, it is worth mentioning that the figures in this paper that correspond to the training phase show the validation losses instead of its training counterparts. This way, we ensure that the model does not suffer from overfitting.

\subsection{Guided Attention}
The most important part of sequence-to-sequence models that use attention is the alignment. The faster the network learns the alignment, the faster the training converges. Therefore, we implemented a guided attention loss where we compare the alignment with a ``straight alignment'' (similar to a multi-variable Gaussian matrix). In our experiments, we identified that it was very effective but if used for too long, it becomes an impediment for the network's training. To tackle this constraint, we introduce the concept of a \textit{guided attention warm-up}. With this strategy, we guide the attention for only \textit{X} steps and then we let the network refine it. Experimentally, we found that the best value for \textit{X} was around 5k. We show a side by side comparison in Figure \ref{fig:alignment} where the speed improvement is noticeable. For visualization purposes, the Y axis has been transposed in both graphs.

\subsection{Model architecture and hyperparameters selection}
To ensure a fair comparison of the models, we conducted experiments to find a common architecture. For this case, we used only the LJ Speech dataset, 88 nodes of input random noise, and the same discriminator with the dimension of each layer equal to 512. We compared two techniques used to help with the training stability, namely WGAN (Wasserstein GAN) and WGAN-GP (WGAN with gradient penalty instead of weight clipping). WGAN proved to be more stable in our experiments. We also used hyperparameters tuning to find the best combination of them.

\subsection{Evaluation strategy}
Evaluating text-to-speech models in an objective way is not trivial. Different datasets will have different losses, different models have different ways of calculating the losses, and training the models takes a very long time. That is why most of the published papers are compared with a subjective Mean Opinion Score (MOS). However, in this paper, MOS or a similar (subjective) metric was not utilized since the main objective of the presented model is not to obtain the highest TTS quality but to tune the generated speech with emotions. Moreover, as far as we know, there is no standard objective metric to evaluate this. 
Hence, an alternative evaluation schema was applied. This strategy relies in a simple idea: six groups of samples are created, each one with a different style token, and for each group we inference 1,000 files. Then, we train a classifier that tries to separate them into those groups, using only generated files. The more the classifier can differentiate them, the better we will consider the GANtron model. The way each group is created differs depending on which type of style token we are using and the objective. When using only noise, each group is assigned a token where each value is random. When using only labels we have five groups with one strong emotion (which value is establish randomly between 0.5 and 1.0) and the rest of the values in the token set to zero. 
Additionally, with the aim of simulating a class with no clear emotional content, a sixth group is also created. In this case, each position of the vector (corresponding to possible emotions) is established with random values form 0 to 1. 
Regarding the case of using the combination of noise and labels, two approaches are evaluated. For this evaluation, an splitting of 85-5-10\% (training-validation-test) was applied. Further details are provided in Section \ref{exp:complete_models}.


\begin{figure*}
    \centering
    \begin{subfigure}[b]{0.45\linewidth}
        \includegraphics[width=\linewidth]{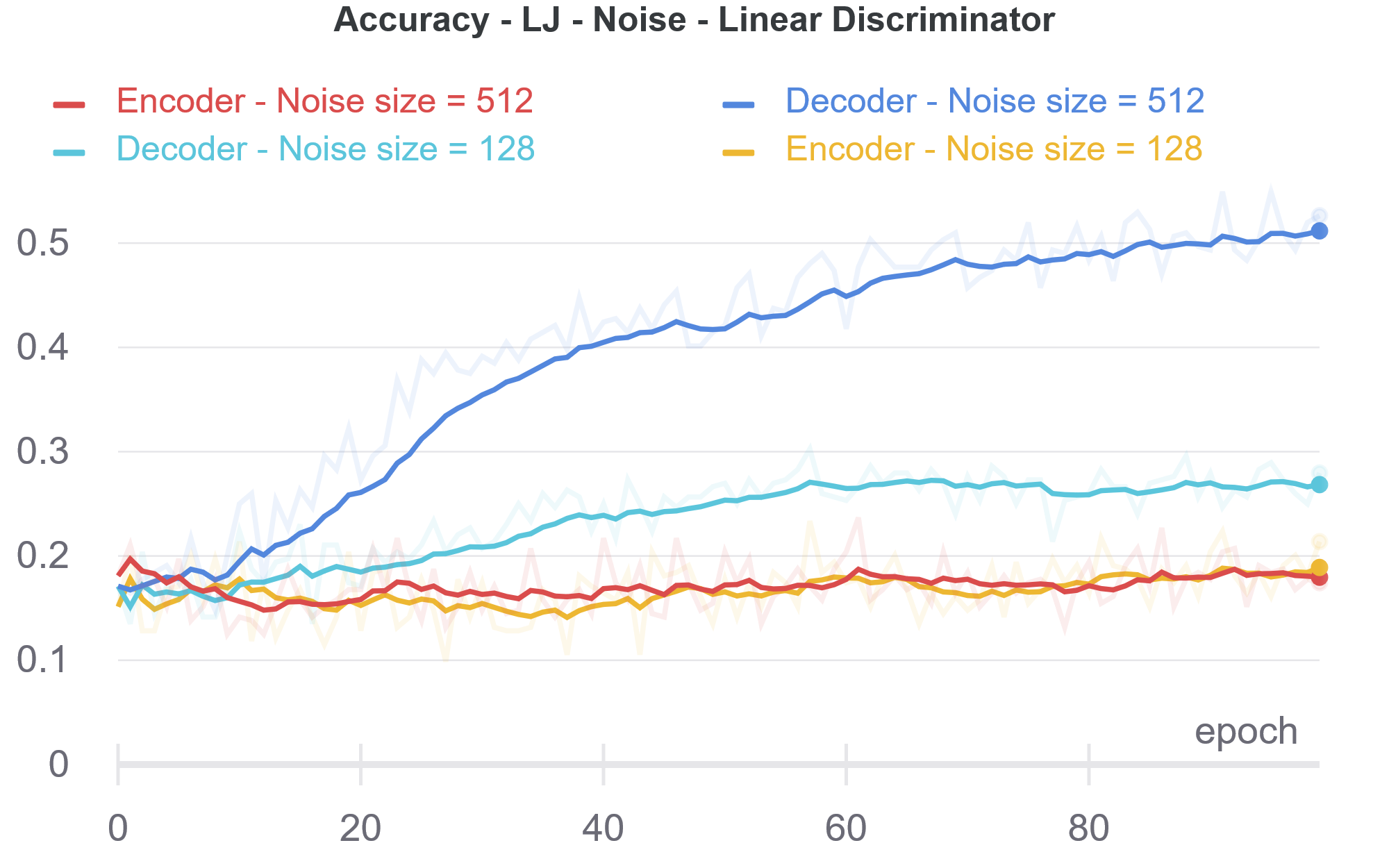}
        \caption{Baseline model: Vanilla GANtron.}
        \label{fig:LJ-noise}
    \end{subfigure}
    \begin{subfigure}[b]{0.45\linewidth}
        \includegraphics[width=\linewidth]{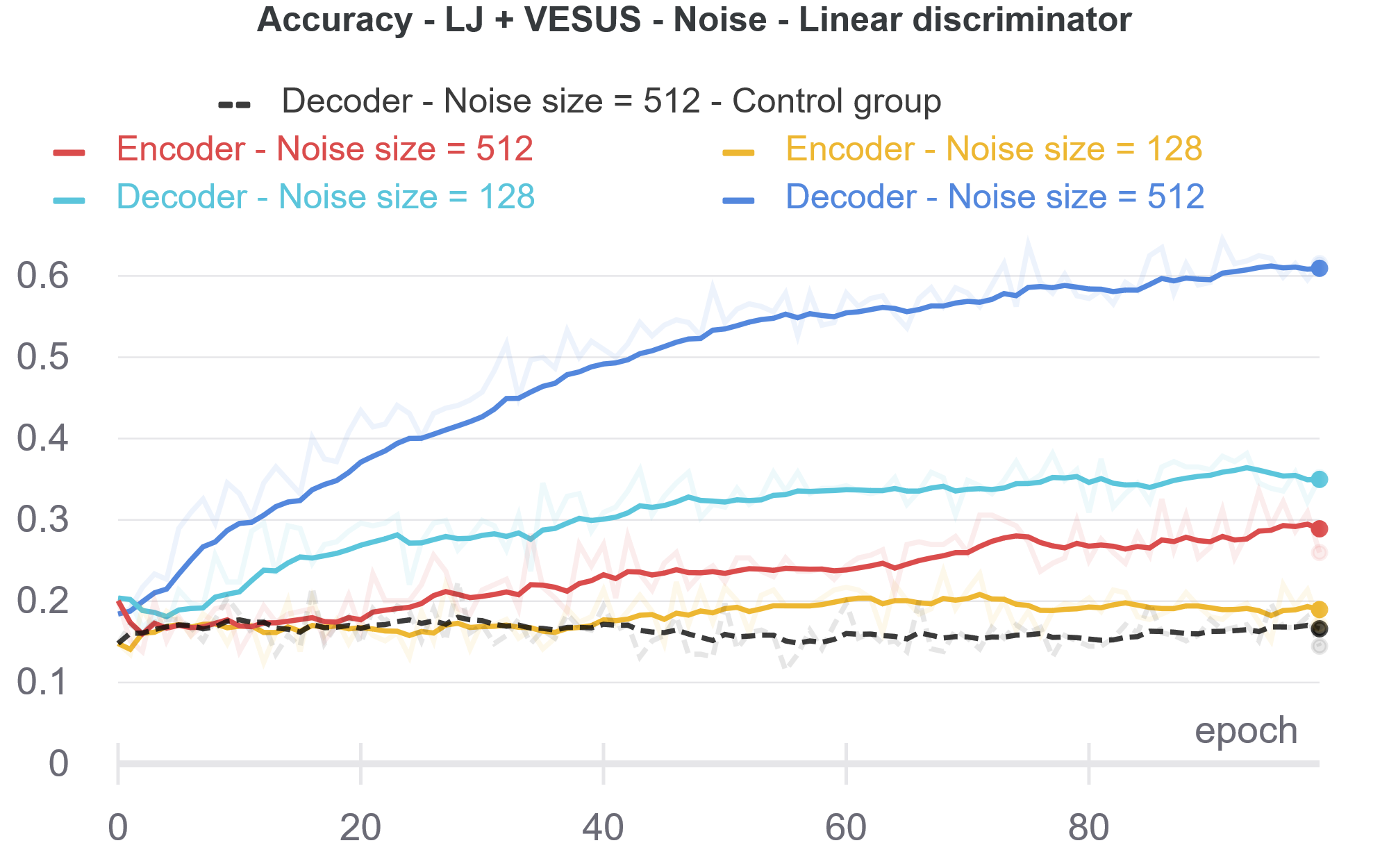}
        \caption{Expressive baseline model (with Linear Discriminator).}
        \label{fig:LJ-VESUS-noise-lD}
    \end{subfigure}
    
    \begin{subfigure}[b]{0.45\linewidth}
        \includegraphics[width=\linewidth]{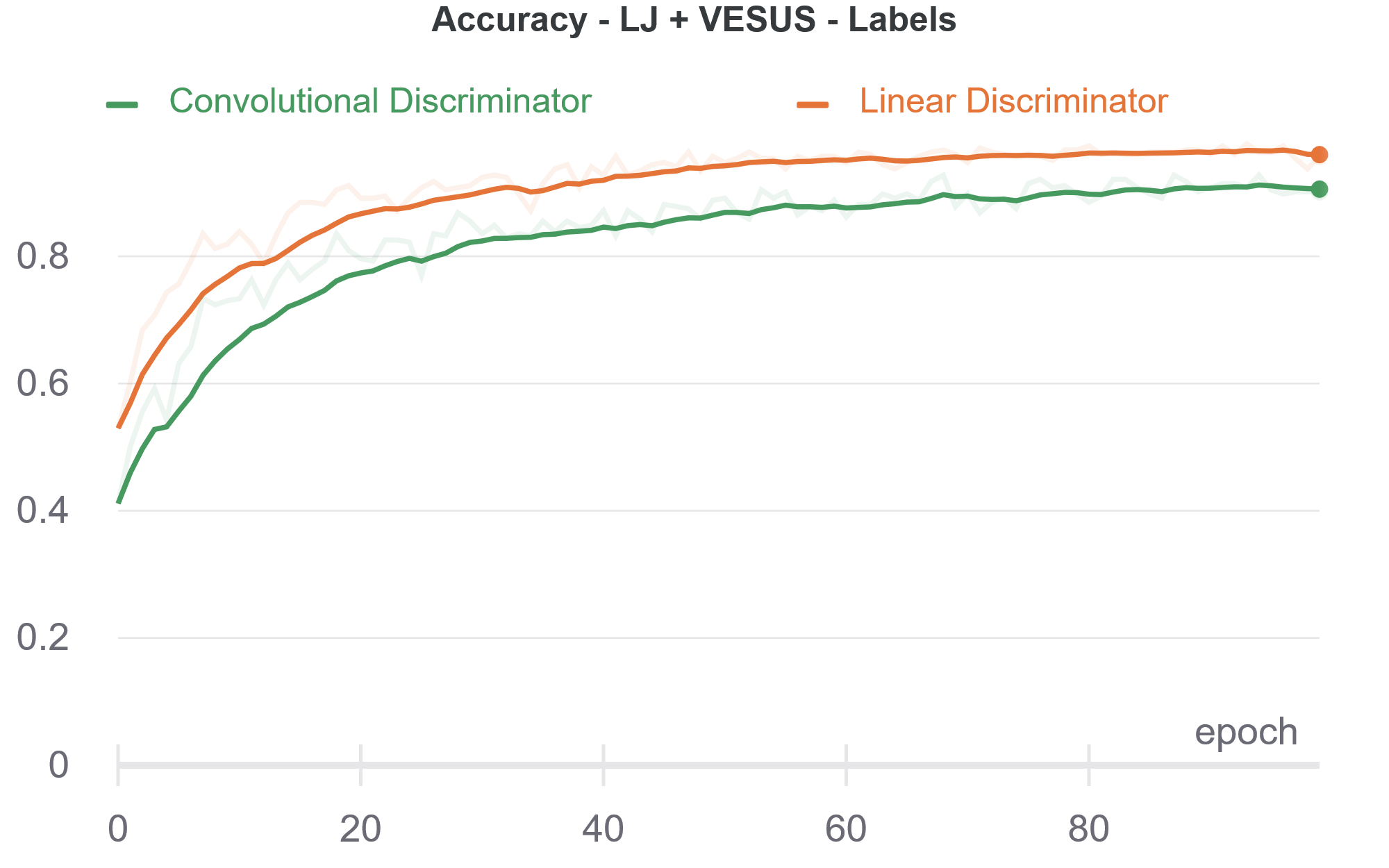}
        \caption{Labelled model.}
        \label{fig:LJ-VESUS-labels}
    \end{subfigure}
    \begin{subfigure}[b]{0.45\linewidth}
        \includegraphics[width=\linewidth]{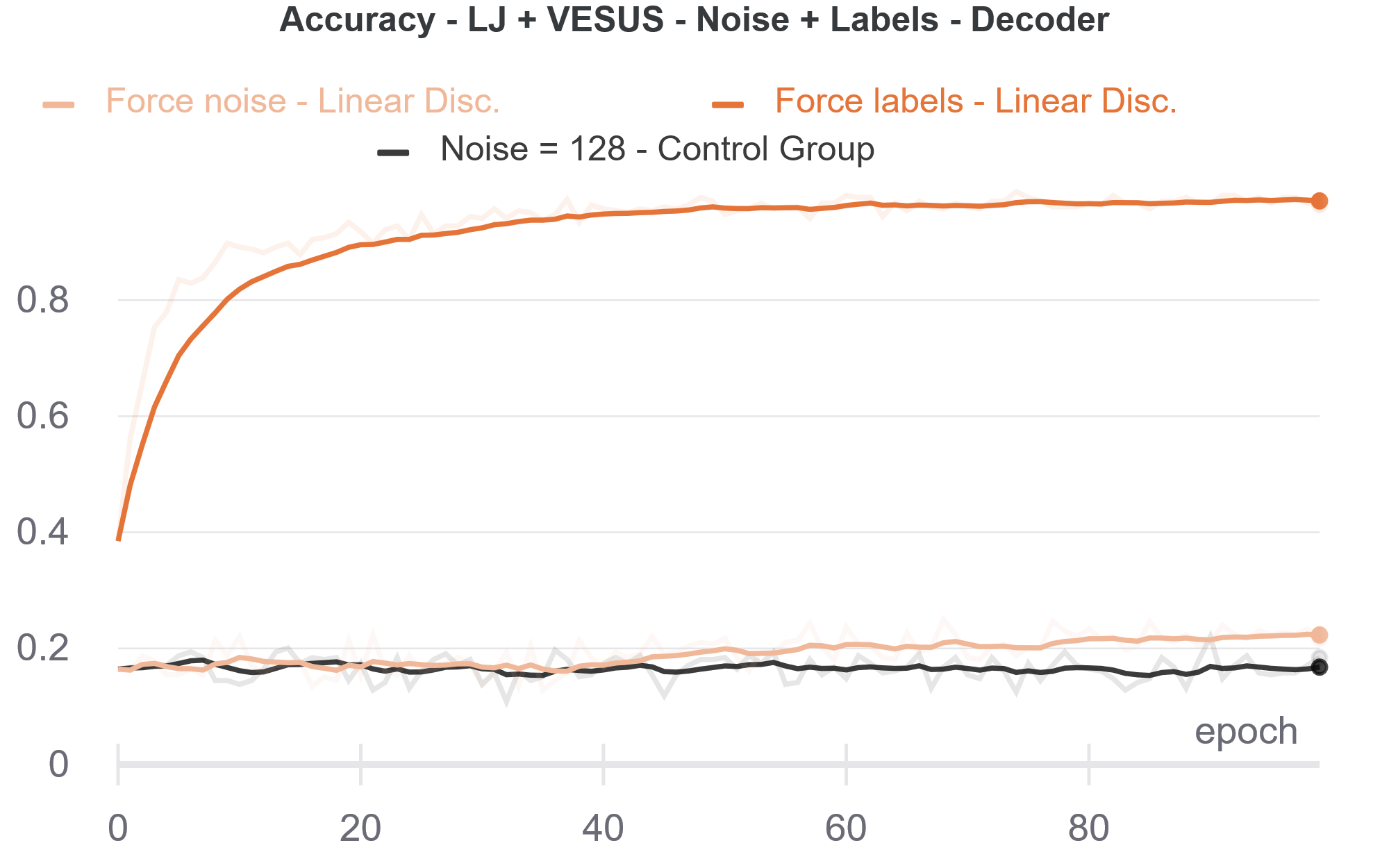}
        \caption{Complete model.}
        \label{fig:LJ-VESUS-noise-labels}
    \end{subfigure}
    \caption{Accuracy of the different models' configurations using, as style token and dataset respectively, only noise and LJ Speech dataset (a), only noise and LJ Speech and VESUS datasets (b), only contextual information and LJ Speech and VESUS datasets (c), and noise and contextual information and LJ Speech and VESUS datasets (d).}
    \label{fig:accuracy-models}
\end{figure*}

\subsection{Baseline model, the Vanilla GANtron}
This is the most basic version of GANtron, where we only use one dataset to train (LJ Speech) and we add noise as style token. In this model, we study the effect that the noise size (128 or 512) and the placement of the style token (input of the encoder or the decoder) have on the performance of the model with the linear discriminator. 

\subsection{Expressive baseline model}
\label{exp:expressive}
The most important difference between VESUS and LJ Speech datasets is that the former has only 253 phrases but in 12,593 files. Therefore, the model will be forced to learn that, for the same utterance, there are many different ways of expressing it. The second difference is that the LJ Speech dataset only has one speaker while VESUS has 10. We hypothesize that the combination of this information helps the model grasp better the emotional content of the audios. Since LJ Speech does not have any annotation regarding the emotions, we set a vector of zeros for the relevant labels. Since we want to test how much can the model generalize the emotions, we test using the LJ speaker.


\subsection{Labelled GANtron model}
\label{exp:labels_only}
As a step toward promoting the emotional content, we take advantage of the labeled dataset VESUS dataset, utilizing this information as style token. In this experiment we compare the performance of the models using the linear and the convolutional discriminators. In this test, we included six groups evaluated as a classification task. The first five groups have one strong emotion (values between 0.5 and 0.8 in the relevant field) and the rest set to 0, the sixth group has random values for each emotion (intended to simulate a class with no clear emotional content).

\subsection{Complete model}
\label{exp:complete_models}
In the previous sections we investigate whether GANtron can generate audio files based on style tokens that were either noise or labels. Therefore, as the next step, we decided to combine them as a single style token. In this experiment we examine not only the accuracy but also, investigate which component (noise or labels) is more important in the new style token. For this purpose we conducted an ablation study applying the same model twice, one where each group has a fixed noise (as in Experiment \ref{exp:expressive}) and the labels of each file are randomized, and another where each group has a fixed label (as in Experiment \ref{exp:labels_only}) and random noise for each file. The results are also compare with a control groups using only noise.

\subsection{Comparison of distributions}\label{sec:comparison}
\begin{figure}
    \centering
    \includegraphics[width=\linewidth]{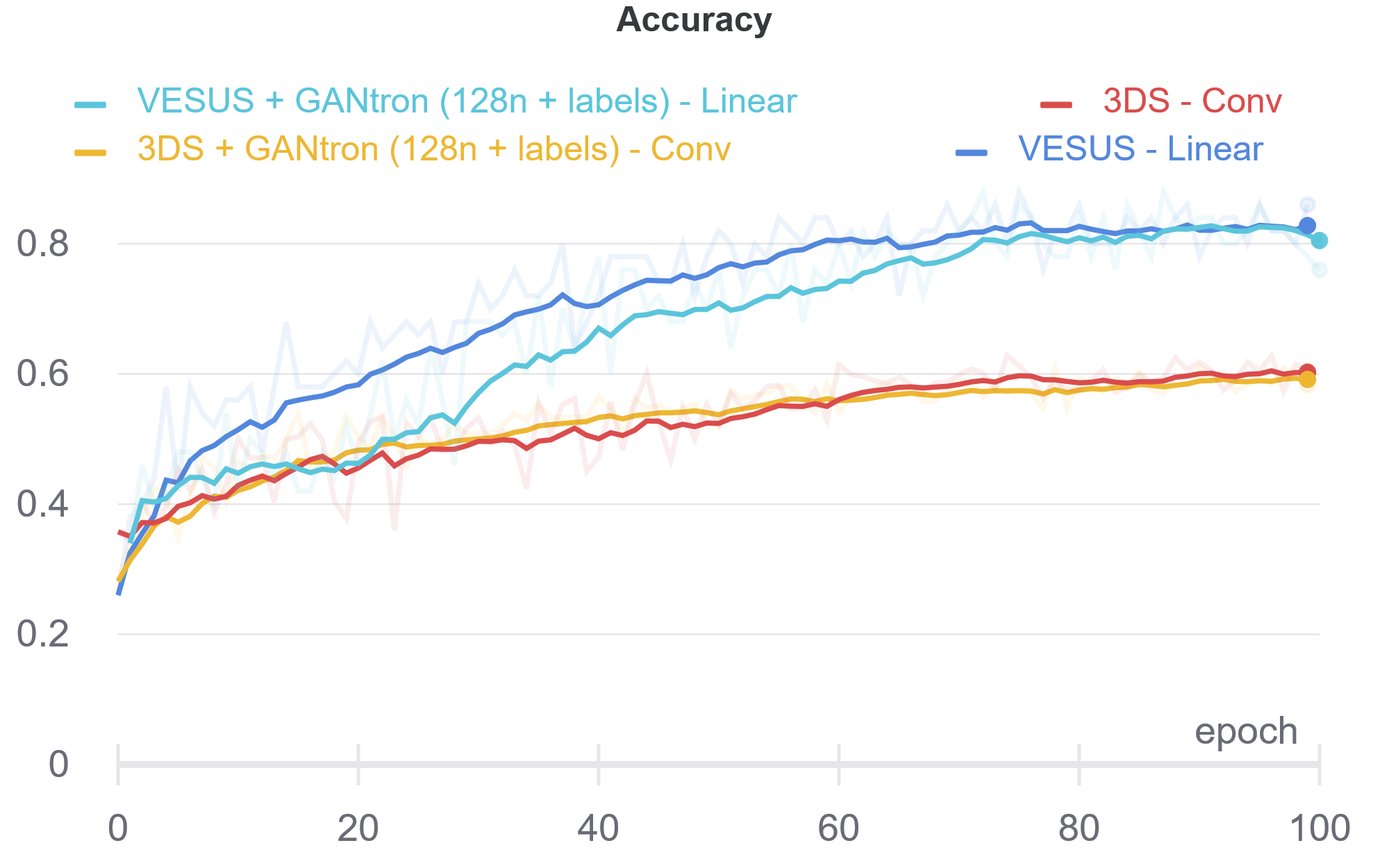}
    \caption{Models trained with GANtron data augmentation vs original data.}
    \label{fig:distribution-check}
\end{figure}

The previous experiments are intended to show whether our models can recreate audios with emotional content. However, we also need to ensure that the generated samples keep the emotional content in the original ones. To evaluate this point, a classifier was trained to differentiate among five emotions, namely anger, fear, happiness, sadness and neutral. This classify is firstly trained with only VESUS data (12,593 files) and subsequently with VESUS and 6,000 generated files using GANtron (increasing the size of the dataset by around 50\%). To classify the audio files, MEL-spectrograms from 80 frames were used. To avoid bias, after every prediction, the starting frame for the 80-frames window is randomized. Additionally, to verify its consistency, we performed the same test using a combination of three datasets, VESUS, CREMA-D and RAVDESS (denoted as 3DS in Figure \ref{fig:distribution-check}). These two new datasets have the same five emotions as VESUS along with some additional ones that, for consistency reasons, are discarded in this experiment.

\section{Results and discussion}
\label{sec:results}
Regarding the Vanilla GANtron model, Figure \ref{fig:LJ-noise} shows the comparison of the four possible combinations (two noise sizes and two possible placements). It is clear that the effect of the style token gets lost if used as input in the encoder. It is also interesting to note that a higher noise dimensionality achieves better performance when used as input of the decoder. The best model (noise size 512, input in the decoder) achieved only a 52\% accuracy. Considering that humans were able to recognize the emotions of the VESUS dataset only with 65\% accuracy and 41\% on CREMA-D, this performance can be considered satisfactory. Nonetheless, listening to the audios we found that the difference between them was far from the intended outcome. 

Figure \ref{fig:LJ-VESUS-noise-lD} shows how using a more expressive dataset in the Vanilla GANtron increases the accuracy by 10 percentage points. We also present a control group where each file uses different random noise and we still assign them to six groups. We can see how it achieves an accuracy of 17\%, what is expected as a random classification, showing that the classifier technique is reliable. 

Both linear and convolutional models achieved very high accuracy (96 and 90\% respectively) on experiment \ref{exp:labels_only} (see Figure \ref{fig:LJ-VESUS-labels}). Unfortunately, 18,03\% of the files created with the convolutional model were erroneous (0\% on the linear model). Meaning that the attention mechanism failed and the inference stopped only because it reached the maximum number of steps allowed. Therefore the use of the convolutional discriminator was discarded.

Figure \ref{fig:LJ-VESUS-noise-labels} shows how the model where the labels were forced with random noise achieves very high accuracy. We conclude that the noise in this model is useful to avoid overfitting. Furthermore, it is not affecting the distribution of the classes of the generated samples. 

We proved that the performance is very similar when having augmentation and not (see Figure \ref{fig:distribution-check}), meaning that the distribution of the GANtron data must be the same as (or very similar to) the original. If the distributions were distanced from each other, the accuracy would drop as it happens when VESUS-based generated samples are used together with CREMA-D and RAVDESS samples (denoted as 3DS). Furthermore, it is important to remember that the speaker in the generated data is different from VESUS and still the performance of the classifier does not drop significantly.

It is worth stressing that the primary objective of this work is to generate reliable text-to-speech outputs with emotional content. As Audio-Emotion Recognition usually suffers from a lack of data, we considered that a data augmentation schema would be appropriate to evaluate whether the generated samples rely on a distribution close to the original samples. However, we do not intend this model as a domain adaption technique but to help with regularization.

\section{Conclusion}
\label{sec:conclusions}
In this paper, we present a novel text-to-spectrogram model pursuing a threefold objective: 1) improve speech synthesis by proposing a new text-to-speech architecture combining the advantages of the Tacotron 2 model and GANs; 2) the incorporation of emotional content into the proposed model to generate more human-like outcomes; and 3) the improvement of the training process with the use of a novel guided attention warm-up that greatly increases the training speed.
Additionally, as intermediate step in an effort to find the most appropriate architecture, several experiments have been conducted considering different parameters such as the kind of input used or certain characteristics of it such as the noise size or the placement to input the emotional information. Finally, different experiments were established to assess the validity of the results obtained.

During this process, each experiment built on the knowledge gathered on the previous ones to test a new hypothesis. We started with a model that used a non-labelled dataset and noise as style token. Then, a second dataset, this time with more emotional content, was used but the style tokens is kept as noise. However, this version achieved better results based on our metrics, proving that GANtron can generalize the emotions with a speaker even when labels are not provided. Furthermore, we showed that we can use noise in the style tokens together with emotions to avoid overfitting. Finally, a model that generates audio files with enough emotional content that can be used for data augmentation of emotion classifiers is proposed. Additionally, for validation purposes, we assess the results of our experiments in a quantitative manner, proving that the generated samples (Mel spectrograms) lie in the same distribution as the original ones.

Some of the main constraints of this and similar works is the data available. Although plenty of dataset have been made accessible in the last years, the challenges related to the annotation of the emotional content represent a limitation when training the algorithms. When exploring five well-known emotional datasets (VESUS, CREMA-D, IEMOCAP, RAVDESS and SAVEE), we identified that they all lack expressiveness and have a dominance of the neutral state with respect to the other emotions. Furthermore, when humans were asked to annotate these datasets and majority voting was applied, around 40 to 60 \% (depending on the dataset) of the annotations had a different emotion than the intended by the actors. For these reasons we believe that the creation of a labeled dataset with high quality and higher intensity ratings would improve the results.

It is worth noting some important future works. Regarding the outcomes of the proposed text-to-speech architecture, even though the quantitative results are promising, the qualitatively assessment of the results would be advisable. 
Furthermore, the research on using adversarial neural networks on text-to-speech synthesis could be expanded. We believe that we have not unlocked the full potential of these networks. A potential modifications could be a change on the loss function used to train the architecture. We would like to explore the possibility of reducing the importance to current loss (a combination of the Wassesstein loss and the loss used by the Tacotron 2 part) using an extra loss. To this effect, the integration of an emotion classifier in the training loop as part of the loss calculation, similar to the work done in other areas with the perceptual loss, especially in the field of computer vision \cite{yin2021} \cite{athanasiadis2020audio}, could lead to a potential improvement.

\bibliographystyle{IEEEtran}
\bibliography{main}


\end{document}